\documentstyle[seceq]{ptptex}

\makeatletter
\newcommand{\ps@persona}{%
\renewcommand{\@oddhead}{\hfil{\makebox{\parbox{2cm}{TOYAMA-103 May 2000}}}}
\renewcommand{\@evenhead}{}
\renewcommand{\@oddfoot}{\hfil\rm{\thepage}\hfil}
\renewcommand{\@evenfoot}{\hfil\rm{\thepage}\hfil}
}
\makeatother

\newcommand{\definition}{\stackrel{\rm d}{\equiv}}

\notypesetlogo  

\title{
Path-Integral Measures in Higher-Derivative Gravities
}

\author{
Shinji {\sc Hamamoto}\footnote{
E-mail: hamamoto@sci.toyama-u.ac.jp} 
\ and Makoto {\sc Nakamura}\footnote{
E-mail: makoto@jodo.sci.toyama-u.ac.jp}
}

\inst{
Department of Physics, Toyama University, \\
Toyama 930-8555, Japan
}

\abst{
A simple method of obtaining path-integral measures in 
higher-derivative gravities is presented. 
The measures are nothing but the generalized Lee-Yang terms.
}

\begin{document}

\maketitle
\thispagestyle{persona}
\pagestyle{plain}

\section{Introduction}

Path-integral measures in higher-derivative gravities have been 
calculated by Buchbinder and Lyahovich.\cite{BL86}
To obtain the results, 
they investigated the structure of the constraints of the theories, 
performing canonical quantization straightforwardly. 

In the present paper, we propose a simple method of deriving 
path-integral measures. 
The measures are obtained by calculating the generalized Lee-Yang 
terms. 

In \S 2 a generic higher-derivative system is considered. 
We review canonical formalism and path-integral quantization 
of the system, showing that the generalized Lee-Yang term gives 
the path-integral measure. 
The Lee-Yang terms are explicitly calculated for Einstein gravity in 
\S 3, 
and for four-derivative gravity in \S 4.
The measures obtained there agree with the ones of 
Buchbinder-Lyahovich. 
The method is also applicable to the case of 
more-than-four-derivative gravities, 
which is studied in \S 5. 
Section 6 gives summary and discussion.

\section{Generalized Lee-Yang term}

We consider a generic system with coordinates 
$x_{a}(t)\ (a=1,\ldots,N)$ 
of Grassmann parities $\epsilon_{a}$.
The Lagrangian of the system is assumed to contain up to 
$n_{a}$-th derivative of $x_{a}(t)$ 
\begin{equation}
L = L(x_{a},\dot{x}_{a},\ddot{x}_{a},\ldots,x^{(n_{a})}_{a}) , 
\label{eqn:201}
\end{equation}
where 
\begin{equation}
x^{(r_{a})}_{a} \definition \left( d/dt\right)^{r_{a}}x_{a}. 
\hspace{3mm}(r_{a}=1,\ldots,n_{a})
\label{eqn:202}
\end{equation}
Canonical formalism of Ostrogradski\cite{O50} regards $x^{(s_{a})}
\ (s_{a}=1,\ldots,n_{a}\! -\! 1)$ as independent 
coordinates $q^{s_{a}+1}$:
\begin{eqnarray}
x^{(s_{a})} & \rightarrow & q^{s_{a}+1},
\label{eqn:203}\\
L(x_{a},\dot{x}_{a},\ldots,x^{(n_{a})}_{a}) 
& \rightarrow & 
L_{\rm q}(q^{1}_{a},\ldots,q^{n_{a}}_{a},\dot{q}^{n_{a}}_{a}) .
\label{eqn:204}
\end{eqnarray}
Momenta conjugate to $q^{n_{a}}_{a}$ are defined as 
usual:\footnote{
In order to distinguish between left and right derivatives, 
we use the following notations:
\[
\begin{array}{lll}
\displaystyle\left(\partial /\partial\theta\right) A &&
{\rm for\ left\ derivative},\\
\displaystyle\partial A/\partial\theta &&
{\rm for\ right\ derivative}.
\end{array}
\]
}
\begin{equation}
p_{n_{a}}^{a}\definition
\partial L_{\rm q}/\partial\dot{q}^{n_{a}}_{a}.
\label{eqn:205}
\end{equation}
The Hessian matrix of $L_{\rm q}$ is 
\begin{equation}
A^{ab}\definition
\left(\partial /\partial\dot{q}^{n_{a}}_{a}\right) 
\partial L_{\rm q}/\partial\dot{q}^{n_{b}}_{b}.
\label{eqn:206}
\end{equation}
If the system is nonsingular ${\rm det}A^{ab}\neq 0$, which is the 
only case considered hereafter, 
then the relation (\ref{eqn:205}) can be inverted to give 
$\dot{q}^{n_{a}}_{a}$ as functions of 
$q^{r_{a}}_{a}\ (r_{a}=1,\ldots,n_{a})$ and $p_{n_{a}}^{a}$:
\begin{equation}
\dot{q}^{n_{a}}_{a}=\dot{q}^{n_{a}}_{a}(q^{r}, p_{n}) .
\label{eqn:207}
\end{equation}
The Hamiltonian is defined by 
\begin{equation}
H\definition
p_{s_{a}}^{a}q^{s_{a}+1}_{a}+
p_{n_{a}}^{a}\dot{q}^{n_{a}}_{a}(q^{r},p_{n}) -
L_{\rm q}(q^{r},\dot{q}^{n}(q^{r},p_{n})) .
\label{eqn:208}
\end{equation}
Canonical equations of motion 
\ 
$\dot{q^{r}}=\left(\partial /\partial p_{r}\right) H$
\ 
and 
\ 
$\dot{p_{r}}=\!\mbox{}-\partial H/\partial q^{r}$
\ 
determine the time development of the system.
Path integral is 
\begin{equation}
Z=
\int{\cal D}q^{r_{a}}_{a}{\cal D}p_{r_{a}}^{a}
\exp\left\{ i\int dt\left[ p_{r_{a}}^{a}\dot{q}^{r_{a}}_{a}-
H(q^{r},p_{r})\right]\right\} .
\label{eqn:209}
\end{equation}
After integration with respect to $p_{s_{a}}^{a}$ and 
$q^{s_{a}+1}_{a}$, 
this reduces to\cite{NH96}
\begin{equation}
Z=
\int{\cal D}q^{1}_{a}{\cal D}p_{n_{a}}^{a}
\exp\left\{ i\int dt\left[ p_{n_{a}}^{a}q^{1(n_{a})}_{a}-
\hat{H}(q^{1},q^{1(s)},p_{n})\right]\right\} ,
\label{eqn:210}
\end{equation}
where 
\begin{eqnarray}
\mbox{}\hspace{-7mm}
\hat{H}(q^{1},q^{1(s)},p_{n}) & \definition &
p_{n_{a}}^{a}\dot{q}^{n_{a}}_{a}(q^{1},q^{1(s)},p_{n}) -
L_{\rm q}(q^{1},q^{1(s)},\dot{q}^{n}(q^{1},q^{1(s)},p_{n})) ,
\label{eqn:211}\\
q^{1(s_{a})}_{a} & \definition &
\left( d/dt\right)^{s_{a}}q^{1}_{a}.
\label{eqn:212}
\end{eqnarray}

To proceed further, take the following special case:
\begin{equation}
L_{\rm q}=\frac{1}{2}\dot{q}^{n_{a}}_{a}A^{ab}\dot{q}^{n_{b}}_{b}+
B^{a}\dot{q}^{n_{a}}_{a}+C,\hspace{3mm}
{\rm det}A^{ab}\neq 0,
\label{eqn:213}
\end{equation}
where $A^{ab}, B^{a}$ and $C$ are arbitrary functions 
of $q^{r_{a}}_{a}$ with the properties
\begin{equation}
\begin{array}{c}
A^{ba}=
(-)^{\epsilon_{a}+\epsilon_{b}+\epsilon_{a}\epsilon_{b}}A^{ab},
\hspace{3mm}
(A^{ab})^{*}=A^{ba},\\
(B^{a})^{*}=(-)^{\epsilon_{a}}B^{a},
\hspace{3mm}
C^{*}=C.
\end{array}
\label{eqn:214}
\end{equation}
In this case, the conjugate momenta are 
\begin{equation}
p_{n_{a}}^{a}=(-)^{\epsilon_{a}}A^{ab}\dot{q}^{n_{b}}_{b}+B^{a}.
\label{eqn:215}
\end{equation}
The Hamiltonian $\hat{H}$ in (\ref{eqn:211}) becomes 
\begin{equation}
\hat{H}=
\frac{1}{2}( p_{n_{a}}^{a}-B^{a}) A_{ab}(-)^{\epsilon_{b}}
( p_{n_{b}}^{b}-B^{b}) -C,
\label{eqn:216}
\end{equation}
where $A_{ab}$ represents the inverse matrix of $A^{ab}$.
Integration with respect to $p_{n_{a}}^{a}$ can be carried out in 
(\ref{eqn:210}).
The result is 
\begin{equation}
Z=
\int{\cal D}q^{1}_{a}\Delta\exp\left( i\int dtL\right).
\label{eqn:217}
\end{equation}
In this expression, $L$ is the higher-derivative Lagrangian in the 
original configuration space 
\begin{equation}
L\definition
\frac{1}{2}q^{1(n_{a})}_{a}A^{ab}q^{1(n_{b})}_{b}+
B^{a}q^{1(n_{a})}_{a}+C.
\label{eqn:218}
\end{equation}
The measure $\Delta$ is given as the generalized Lee-Yang term 
\begin{equation}
\Delta\definition
\left({\rm det}A_{ab}\right)^{-1/2}=
\left({\rm det}A^{ab}\right)^{1/2}.
\label{eqn:219}
\end{equation}

\section{Measure in Einstein gravity}

The Lagrangian is given by
\begin{eqnarray}
{\cal L}^{(1)} & = & 
{\cal L}_{\rm E}+{\cal L}_{\rm BRS}^{(1)},
\label{eqn:301}\\
{\cal L}_{\rm E} & = & 
\mbox{}-\frac{1}{\kappa^{2}}\sqrt{-g}R. 
\label{eqn:302}
\end{eqnarray}
For the BRS Lagrangian we take
\begin{eqnarray}
{\cal L}_{\rm BRS}^{(1)} & = & 
\mbox{}-i\delta\left[
\bar{c}_{\mu}\left(\frac{1}{\kappa}\partial_{\nu}\tilde{g}^{\mu\nu}
-\frac{\alpha}{2}\eta^{\mu\nu}b_{\nu}\right)\right]
\nonumber\\
& = & {\cal L}_{\rm GF}^{(1)}+{\cal L}_{\rm FP}^{(1)},
\label{eqn:303}\\
{\cal L}_{\rm GF}^{(1)} & = & 
\frac{1}{\kappa}b_{\mu}\partial_{\nu}\tilde{g}^{\mu\nu}
-\frac{\alpha}{2}\eta^{\mu\nu}b_{\mu}b_{\nu},
\label{eqn:304}\\
{\cal L}_{\rm FP}^{(1)} & = & 
\mbox{}-\frac{i}{2}
(\partial_{\mu}\bar{c}_{\nu}+\partial_{\nu}\bar{c}_{\mu})
D^{\mu\nu}_{\rho}c^{\rho}.
\label{eqn:305}
\end{eqnarray}
Here $\kappa$ is the gravitational constant; 
$\alpha$ is a gauge parameter; 
$\eta_{\mu\nu}$ is the Minkowski metric $(-\! +\! ++)$; 
$c^{\mu},\bar{c}_{\mu}$ are the Faddeev-Popov (FP) ghosts; 
$b_{\mu}$ are the Nakanishi-Lautrap (NL) fields; 
$\tilde{g}^{\mu\nu}\definition \sqrt{-g}g^{\mu\nu}$; 
and
\begin{equation}
D^{\mu\nu}_{\rho}\definition 
\tilde{g}^{\mu\sigma}\delta^{\nu}_{\rho}\partial_{\sigma}+
\tilde{g}^{\nu\sigma}\delta^{\mu}_{\rho}\partial_{\sigma}-
\tilde{g}^{\mu\nu}\partial_{\rho}-
(\partial_{\rho}\tilde{g}^{\mu\nu}).
\label{eqn:306}
\end{equation}
The Lagrangian ${\cal L}^{(1)}$ is invariant 
under the following BRS transformation:
\begin{equation}
\left\{
\begin{array}{rcl}
\delta\tilde{g}^{\mu\nu} & = & 
\kappa D^{\mu\nu}_{\rho}c^{\rho},\\
\delta c^{\mu} & = & 
\mbox{}-\kappa c^{\lambda}\partial_{\lambda}c^{\mu},\\
\delta\bar{c}_{\mu} & = & ib_{\mu},\\
\delta b_{\mu} & = & 0.
\end{array}
\right.
\label{eqn:307}
\end{equation}
Because the BRS Lagrangian ${\cal L}_{\rm BRS}^{(1)}$ is introduced 
from the beginning, the system is made nonsingular.
In order to calculate the Hessian matrix it is convenient to 
rewrite ${\cal L}_{\rm GF}^{(1)}$ by performing 
path-integration with respect to $b_{\mu}$ in advance:
\begin{equation}
{\cal L}_{\rm GF}^{(1)} \longrightarrow 
{\cal L}_{\rm GF}^{(1)'} = 
\frac{1}{2\alpha\kappa^{2}}\eta_{\mu\nu}
\partial_{\rho}\tilde{g}^{\mu\rho}
\partial_{\sigma}\tilde{g}^{\nu\sigma}.
\label{eqn:308}
\end{equation}

Let the ADM coordinates be $(\lambda, \lambda_{i}, e_{ij})$:
\begin{equation}
g_{\mu\nu}=
\left(
 \begin{array}{cc}
 \lambda^{k}\lambda_{k}-\lambda^{2} & \lambda_{j} \\
 \lambda_{i}                        & e_{ij}
 \end{array}
\right), 
\hspace{3mm}\lambda^{i}\definition e^{ij}\lambda_{j}.
\label{eqn:309}
\end{equation}
Expressing the Lagrangians 
(\ref{eqn:302}),(\ref{eqn:308}) and (\ref{eqn:305}) 
in terms of the ADM coordinates, 
we have 
\begin{eqnarray}
{\cal L}_{\rm E} & = & 
\frac{1}{\kappa^{2}}\frac{1}{4\lambda}\sqrt{e}
\left(\mbox{}-P^{(2)ij,kl}+2P^{(0)ij,kl}\right)
\dot{e}_{ij}\dot{e}_{kl}+
\ldots , 
\label{eqn:310}\\
{\cal L}_{\rm GF}^{(1)'} & = & 
\frac{1}{2\alpha\kappa^{2}}e
\left\{
\mbox{}-\frac{1}{\lambda^{4}}
\left( 1-\eta_{mn}\lambda^{m}\lambda^{n}\right)
\dot{\lambda}^{2}
-\frac{2}{\lambda^{3}}\eta_{mn}\lambda^{m}e^{ni}
\dot{\lambda}\dot{\lambda}_{i}\right.
\nonumber\\ & &
\mbox{}+\frac{1}{\lambda^{3}}
\left[
\left( 1-\eta_{mn}\lambda^{m}\lambda^{n}\right) e^{ij}
+\eta_{mn}\lambda^{m}
\left( e^{ni}\lambda^{j}+e^{nj}\lambda^{i}\right)
\right]
\dot{\lambda}\dot{e}_{ij}
\nonumber\\ & &
\mbox{}+\frac{1}{\lambda^{2}}\eta_{mn}e^{mi}e^{nj}
\dot{\lambda}_{i}\dot{\lambda}_{j}
+\frac{1}{\lambda^{2}}\eta_{mn}e^{mi}
\left(
\lambda^{n}e^{kl}-e^{nk}\lambda^{l}-e^{nl}\lambda^{k}
\right)
\dot{\lambda}_{i}\dot{e}_{kl}
\nonumber\\ & &
\mbox{}+\frac{1}{4\lambda^{2}}
\left[
\mbox{}-e^{ij}e^{kl}\right.
\nonumber\\ & &
\left.\!
\mbox{}+\eta_{mn}
\left(
\lambda^{m}e^{ij}-e^{mi}\lambda^{j}-e^{mj}\lambda^{i}
\right)\left(
\lambda^{n}e^{kl}-e^{nk}\lambda^{l}-e^{nl}\lambda^{k}
\right)
\right]
\dot{e}_{ij}\dot{e}_{kl}
\left.\frac{}{}\!\!\!\right\}
\nonumber\\ & &
\mbox{}+\ldots ,
\label{eqn:311}\\
{\cal L}_{\rm FP}^{(1)} & = & 
i\sqrt{e}
\left\{
\frac{1}{\lambda}\dot{\bar{c}}_{0}\dot{c}^{0}
+\frac{1}{\lambda^{2}}\dot{\bar{c}}_{0}c^{0}\dot{\lambda}
-\frac{1}{2\lambda}\dot{\bar{c}}_{0}c^{0}e^{ij}\dot{e}_{ij}
+\frac{1}{\lambda}\dot{\bar{c}}_{i}\dot{c}^{i}\right.
\nonumber\\ & &
\mbox{}+\frac{1}{\lambda}\dot{\bar{c}}_{i}c^{0}
\left[
\mbox{}-\frac{1}{\lambda}e^{ij}\lambda_{j}\dot{\lambda}
+e^{ij}\dot{\lambda}_{j}
+\frac{1}{2}\left(
e^{ij}e^{kl}-e^{ik}e^{jl}-e^{il}e^{jk}\right)
\lambda_{j}\dot{e}_{kl}
\right]
\left.\frac{}{}\!\!\!\right\}
\nonumber\\ & &
\mbox{}+\ldots ,
\label{eqn:312}
\end{eqnarray}
where the projection operators 
\begin{equation}
P^{(0)ij,kl}\definition\frac{1}{3}e^{ij}e^{kl}, \makebox[3mm]{}
P^{(2)ij,kl}\definition\frac{1}{2}\left( 
e^{ik}e^{jl}+e^{il}e^{jk}\right) -
\frac{1}{3}e^{ij}e^{kl} 
\label{eqn:313} 
\end{equation}
have been introduced. 
In (\ref{eqn:310})-(\ref{eqn:312}),
only the highest-derivative terms have been written down explicitly. 

These terms are sufficient to read the Hessian matrix 
\begin{equation}
M^{(1)} = 
\begin{array}{r@{}l}
  & \begin{array}{ccccccc}
      \makebox[0.7em]{}e_{kl} & 
      \makebox[-0.5em]{}\lambda & 
      \makebox[-0.5em]{}\lambda_{b} & 
      \makebox[-0.5em]{}c^{0} & 
      \makebox[-0.5em]{}\bar{c}_{0} & 
      \makebox[-0.5em]{}c^{j} & 
      \makebox[-0.5em]{}\bar{c}_{j} 
    \end{array}\\
  \begin{array}{c}
    e_{ij} \\ \lambda \\ \lambda_{a} \\ c^{0} \\ \bar{c}_{0} \\
    c^{i} \\ \bar{c}_{i} 
  \end{array}
  & \left( 
      \begin{minipage}{40mm}
      \begin{tabular}{ccc|cccc}
        & &\makebox[-0.6em]{} & &          & 
        \makebox[-0.8em]{} & \\
        &$A^{(1)}$ &\makebox[-0.6em]{} & & 
        \makebox[0.5em]{}$\Gamma^{(1)}$ & 
        \makebox[-0.8em]{} & \\
        & &\makebox[-0.6em]{} & &          & 
        \makebox[-0.8em]{} & \\ \hline
        & &\makebox[-0.6em]{} & &     & 
        \makebox[-0.8em]{} & \\
        & &\makebox[-0.6em]{} & &     & 
        \makebox[-0.8em]{} & \\
        & \raisebox{1.5ex}{$\Gamma^{(1)t}$} &\makebox[-0.6em]{} & & 
        \makebox[0.5em]{}\raisebox{1.5ex}{$B^{(1)}$} & 
        \makebox[-0.8em]{} & \\
        & &\makebox[-0.6em]{} & &     & 
        \makebox[-0.8em]{} & 
      \end{tabular}
      \end{minipage}
    \right) .
\end{array}
\label{eqn:314}
\end{equation}
The submatrix $A^{(1)}$
\begin{equation}
A^{(1)}=\left(
 \begin{array}{ccc}
  A^{(1)ij,kl} & A^{(1)ij,0} & A^{(1)ij,b} \\
  A^{(1)0,kl} & A^{(1)00} & A^{(1)0b} \\
  A^{(1)a,kl} & A^{(1)a0} & A^{(1)ab} 
 \end{array}
\right)
\label{eqn:315}
\end{equation}
has the elements 
\begin{eqnarray}
\mbox{}\hspace{-7mm}
A^{(1)ij,kl} & = & 
\frac{1}{\kappa^{2}}\frac{\sqrt{e}}{2\lambda}
\left(\mbox{}-P^{(2)ij,kl}+2P^{(0)ij,kl}\right)
\nonumber\\ & &
\mbox{}+\frac{1}{2\alpha\kappa^{2}}\frac{e}{2\lambda^{2}}
\left[
\mbox{}-e^{ij}e^{kl}\right.
\nonumber\\ & &
\left.\!
\mbox{}+\eta_{mn}
\left(
\lambda^{m}e^{ij}-e^{mi}\lambda^{j}-e^{mj}\lambda^{i}
\right)\left(
\lambda^{n}e^{kl}-e^{nk}\lambda^{l}-e^{nl}\lambda^{k}
\right)
\right] ,
\label{eqn:316}\\
\mbox{}\hspace{-7mm}
A^{(1)ij,0} & = & 
\frac{1}{2\alpha\kappa^{2}}\frac{e}{\lambda^{3}}
\left[
\left( 1-\eta_{mn}\lambda^{m}\lambda^{n}\right) e^{ij}
+\eta_{mn}\lambda^{m}
\left( e^{ni}\lambda^{j}+e^{nj}\lambda^{i}\right)
\right] ,
\label{eqn:317}\\
\mbox{}\hspace{-7mm}
A^{(1)ij,b} & = & 
\frac{1}{2\alpha\kappa^{2}}\frac{e}{\lambda^{2}}
\eta_{mn}e^{mi}
\left(
\lambda^{n}e^{ij}-e^{ni}\lambda^{j}-e^{nj}\lambda^{i}
\right) ,
\label{eqn:318}\\
\mbox{}\hspace{-7mm}
A^{(1)00} & = & 
\frac{1}{2\alpha\kappa^{2}}
\left( -\frac{2e}{\lambda^{4}}\right)
\left( 1-\eta_{mn}\lambda^{m}\lambda^{n}\right) ,
\label{eqn:319}\\
\mbox{}\hspace{-7mm}
A^{(1)0b} & = & 
\frac{1}{2\alpha\kappa^{2}}
\left( -\frac{2e}{\lambda^{3}}\right)
\eta_{mn}\lambda^{m}e^{nb},
\label{eqn:320}\\
\mbox{}\hspace{-7mm}
A^{(1)ab} & = & 
\frac{1}{2\alpha\kappa^{2}}
\frac{2e}{\lambda^{2}}\eta_{mn}e^{ma}e^{nb}.
\label{eqn:321}
\end{eqnarray}
The submatrix $B^{(1)}$ has the form 
\begin{equation}
B^{(1)}=\left(
 \begin{array}{cccc}
 J^{(1)} &   &   & \\
   & J^{(1)} &   & \\
   &   & J^{(1)} & \\
   &   &   & J^{(1)} 
 \end{array}
\right) ,\makebox[3mm]{}
J^{(1)}=i\frac{\sqrt{e}}{\lambda}
\left(
 \begin{array}{cc}
   & -1 \\
 1 & 
 \end{array}
\right) .
\label{eqn:322}
\end{equation}

It is easy to calculate the determinants of these submatrices:
\begin{eqnarray}
\det A^{(1)} & \propto & \lambda^{-16}e,
\label{eqn:323}\\
\det B^{(1)} & \propto & \lambda^{-8}e^{4}.
\label{eqn:324}
\end{eqnarray}
Using the fact $\Gamma^{(1)}B^{(1)-1}\Gamma^{(1)t}=0$, we have
\begin{equation}
\det M^{(1)}=
\frac{\det (A^{(1)}-\Gamma^{(1)}B^{(1)-1}\Gamma^{(1)t})}
{\det B^{(1)}}= 
\frac{\det A^{(1)}}{\det B^{(1)}}\propto
\lambda^{-8}e^{-3}.
\label{eqn:325}
\end{equation}
Changing field variables from the ADM 
coordinates $(\lambda,\lambda_{i},e_{ij})$ 
to the original ones $g_{\mu\nu}$ gives the factor 
\begin{eqnarray}
(g^{00})^{m}(-g)^{n}{\cal D}g_{\mu\nu} & = & 
\lambda^{-2m+2n+1}e^{n}{\cal D}\lambda{\cal D}\lambda_{i}
{\cal D}e_{ij} \nonumber\\
& = & 
(\det M^{(1)})^{1/2}{\cal D}\lambda{\cal D}\lambda_{i}
{\cal D}e_{ij}.
\label{eqn:326}
\end{eqnarray}
In the present case Eq.(\ref{eqn:325}) 
shows $m=1$ and $n=-3/2$.
Thus the path-integral 
measure $\Delta^{(1)}$ is
\begin{equation}
\Delta^{(1)} = g^{00}(-g)^{-3/2}.
\label{eqn:327}
\end{equation}
Bringing back the $b_{\mu}$-integration 
returns the gauge-fixing Lagrangian 
${\cal L}_{\rm GF}^{(1)'}$ 
to the original form 
${\cal L}_{\rm GF}^{(1)}$. 
We finally obtain 
\begin{equation}
Z^{(1)}=
\int{\cal D}g_{\mu\nu}{\cal D}b_{\mu}{\cal D}c^{\mu}
{\cal D}\bar{c}_{\mu}\Delta^{(1)}\exp\left(
i\int{\cal L}^{(1)}d^{4}x\right) .
\label{eqn:328}
\end{equation}

\section{Measure in four-derivative gravity}

The Lagrangian of four-derivative gravity is given by 
\begin{eqnarray}
{\cal L}^{(2)} & = & 
{\cal L}_{\rm E}+{\cal L}_{\rm HD}^{(2)}+{\cal L}_{\rm BRS}^{(2)},
\label{eqn:401}\\
{\cal L}_{\rm HD}^{(2)} & = & 
\sqrt{-g}\left( \beta_{2}R^{\mu\nu}R_{\mu\nu}+
\beta_{0}R^{2}\right),
\label{eqn:402}\\
{\cal L}_{\rm BRS}^{(2)} & = & 
{\cal L}_{\rm BRS}^{(1)} 
\label{eqn:403}
\end{eqnarray}
with the parameters of the theory $\beta_{2},\beta_{0}$.
For the BRS Lagrangian we have adopted the same one as in the 
previous section.
This is sufficient for making the theory nonsingular.

By the use of the ADM coordinates 
the Lagrangians are expressed as 
\begin{eqnarray}
{\cal L}_{\rm E}+
{\cal L}_{\rm HD}^{(2)}
& = & 
\frac{1}{4\lambda^{3}}\sqrt{e}
\Lambda^{(2)ij,kl}
\nonumber\\
& & 
\makebox[2em]{}\times\left[
\ddot{e}_{ij}-\frac{\dot{\lambda}}{\lambda}\left(
\dot{e}_{ij}-\nabla_{i}\lambda_{j}-\nabla_{j}\lambda_{i}\right)
-\left(\nabla_{i}\dot{\lambda}_{j}+\nabla_{j}\dot{\lambda}_{i}
\right)\right]
\nonumber\\
& & 
\makebox[2em]{}\times\left[
\ddot{e}_{kl}-\frac{\dot{\lambda}}{\lambda}\left(
\dot{e}_{kl}-\nabla_{k}\lambda_{l}-\nabla_{l}\lambda_{k}\right)
-\left(\nabla_{k}\dot{\lambda}_{l}+\nabla_{l}\dot{\lambda}_{k}
\right)\right]
\nonumber\\
& & 
\mbox{}+\ldots ,
\label{eqn:404}\\
{\cal L}_{\rm GF}^{(2)'} & = & {\cal L}_{\rm GF}^{(1)'}
\nonumber\\
& = & 
\frac{1}{2\alpha\kappa^{2}}e\left[
\mbox{}-\frac{1}{\lambda^{4}}\left(
1-\eta_{mn}\lambda^{m}\lambda^{n}\right)\dot{\lambda}^{2}\right.
\nonumber\\
& & 
\left.\makebox[4em]{}\!\!\!
-\frac{2}{\lambda^{3}}\eta_{mn}\lambda^{m}e^{ni}
\dot{\lambda}\dot{\lambda}_{i}
+\frac{1}{\lambda^{2}}\eta_{mn}e^{mi}e^{nj}
\dot{\lambda}_{i}\dot{\lambda}_{j}
\right]
\nonumber\\
& & 
\mbox{}+\ldots ,
\label{eqn:405}\\
{\cal L}_{\rm FP}^{(2)}& = & {\cal L}_{\rm FP}^{(1)}
\nonumber\\
& = & 
i\sqrt{e}\left[
\frac{1}{\lambda}\dot{\bar{c}}_{0}\dot{c}^{0}
+\frac{1}{\lambda^{2}}\dot{\bar{c}}_{0}c^{0}\dot{\lambda}
+\frac{1}{\lambda}\dot{\bar{c}}_{i}\dot{c}^{i}
+\frac{1}{\lambda}\dot{\bar{c}}_{i}c^{0}\left(
\mbox{}-\frac{1}{\lambda}\lambda^{i}\dot{\lambda}
+e^{ij}\dot{\lambda}_{j}\right)\right]
\nonumber\\
& & 
\mbox{}+\ldots ,
\label{eqn:406}
\end{eqnarray}
where $\Lambda^{(2)ij,kl}$ represents 
\begin{equation}
\Lambda^{(2)ij,kl}\definition
\beta_{2}P^{(2)ij,kl}+
4\left( \beta_{2}+3\beta_{0}\right) P^{(0)ij,kl},
\label{eqn:407}
\end{equation}
and $\nabla_{i}$ denotes the covariant derivative associated with
$e_{ij}$.
We assume $\beta_{2}\neq 0$ and $\beta_{2}+3\beta_{0}\neq 0$ 
from now on.
In this case the method of obtaining path-integral measures
explained in \S 2
is applicable without any complexity.

The Hessian matrix
\begin{equation}
M^{(2)} = 
\begin{array}{r@{}l}
  & \begin{array}{ccccccc}
      \makebox[0.7em]{}e_{kl} & 
      \makebox[-0.5em]{}\lambda & 
      \makebox[-0.5em]{}\lambda_{b} & 
      \makebox[-0.5em]{}c^{0} & 
      \makebox[-0.5em]{}\bar{c}_{0} & 
      \makebox[-0.5em]{}c^{j} & 
      \makebox[-0.5em]{}\bar{c}_{j} 
    \end{array}\\
  \begin{array}{c}
    e_{ij} \\ \lambda \\ \lambda_{a} \\ c^{0} \\ \bar{c}_{0} \\
    c^{i} \\ \bar{c}_{i} 
  \end{array}
  & \left( 
      \begin{minipage}{40mm}
      \begin{tabular}{ccc|cccc}
        & &\makebox[-0.6em]{} & &          & 
        \makebox[-0.8em]{} & \\
        &$A^{(2)}$ &\makebox[-0.6em]{} & & 
        \makebox[0.5em]{}$\Gamma^{(2)}$ & 
        \makebox[-0.8em]{} & \\
        & &\makebox[-0.6em]{} & &          & 
        \makebox[-0.8em]{} & \\ \hline
        & &\makebox[-0.6em]{} & &     & 
        \makebox[-0.8em]{} & \\
        & &\makebox[-0.6em]{} & &     & 
        \makebox[-0.8em]{} & \\
        & \raisebox{1.5ex}{$\Gamma^{(2)t}$} &\makebox[-0.6em]{} & & 
        \makebox[0.5em]{}\raisebox{1.5ex}{$B^{(2)}$} & 
        \makebox[-0.8em]{} & \\
        & &\makebox[-0.6em]{} & &     & 
        \makebox[-0.8em]{} & 
      \end{tabular}
      \end{minipage}
    \right)
\end{array}
\label{eqn:408}
\end{equation}
can be read from (\ref{eqn:404})-(\ref{eqn:406}) 
as follows:
\begin{equation}
A^{(2)}=\left(
 \begin{array}{ccc}
  A^{(2)ij,kl} & A^{(2)ij,0} & A^{(2)ij,b} \\
  A^{(2)0,kl} & A^{(2)00} & A^{(2)0b} \\
  A^{(2)a,kl} & A^{(2)a0} & A^{(2)ab} 
 \end{array}
\right)
\label{eqn:409}
\end{equation}
has the elements 
\begin{eqnarray}
\mbox{}\hspace{-5mm}
A^{(2)ij,kl} & = & 
\frac{\sqrt{e}}{2\lambda^{3}}
\Lambda^{(2)ij,kl},
\label{eqn:410}\\
\mbox{}\hspace{-5mm}
A^{(2)ij,0} & = & 
\mbox{}-\frac{\sqrt{e}}{2\lambda^{4}}
\Lambda^{(2)ij,kl}
\left( \dot{e}_{kl}-\nabla_{k}\lambda_{l}-\nabla_{l}\lambda_{k}
\right) ,
\label{eqn:411}\\
\mbox{}\hspace{-5mm}
A^{(2)ij,b} & = & 
\mbox{}-\frac{\sqrt{e}}{2\lambda^{3}}
\Lambda^{(2)ij,kl}
\left( \nabla_{k}\delta_{l}^{b}+\nabla_{l}\delta_{k}^{b}
\right) ,
\label{eqn:412}\\
\mbox{}\hspace{-5mm}
A^{(2)00} & = & 
\frac{1}{2\alpha\kappa^{2}}
\left( -\frac{2e}{\lambda^{4}}\right)
\left( 1-\eta_{mn}\lambda^{m}\lambda^{n}\right)
\nonumber\\
& & 
\mbox{}+\frac{\sqrt{e}}{2\lambda^{5}}
\Lambda^{(2)ij,kl}
\left( \dot{e}_{ij}-\nabla_{i}\lambda_{j}-\nabla_{j}\lambda_{i}
\right)
\left( \dot{e}_{kl}-\nabla_{k}\lambda_{l}-\nabla_{l}\lambda_{k}
\right) ,
\label{eqn:413}\\
\mbox{}\hspace{-5mm}
A^{(2)0b} & = & 
\frac{1}{2\alpha\kappa^{2}}
\left( -\frac{2e}{\lambda^{3}}\right)
\eta_{mn}\lambda^{m}e^{nb}
\nonumber\\
& & 
\mbox{}+\frac{\sqrt{e}}{2\lambda^{4}}
\Lambda^{(2)ij,kl}
\left( \dot{e}_{ij}-\nabla_{i}\lambda_{j}-\nabla_{j}\lambda_{i}
\right)
\left( \nabla_{k}\delta_{l}^{b}+\nabla_{l}\delta_{k}^{b}\right) ,
\label{eqn:414}\\
\mbox{}\hspace{-5mm}
A^{(2)ab} & = & 
\frac{1}{2\alpha\kappa^{2}}
\frac{2e}{\lambda^{2}}
\eta_{mn}e^{ma}e^{nb}
\nonumber\\
& & 
\mbox{}+\left( 
\stackrel{\leftarrow}{\nabla}_{i}\!\delta_{j}^{a}
+\stackrel{\leftarrow}{\nabla}_{j}\!\delta_{i}^{a}
\right)\frac{\sqrt{e}}{2\lambda^{3}}
\Lambda^{(2)ij,kl}
\left( \nabla_{k}\delta_{l}^{b}+\nabla_{l}\delta_{k}^{b}\right) ,
\label{eqn:415}
\end{eqnarray}
and $B^{(2)}$ is
\begin{equation}
B^{(2)}=B^{(1)}.
\label{eqn:416}
\end{equation}

The determinant of $A^{(2)}$ can be calculated in this case as 
\begin{equation}
\det A^{(2)}\propto \lambda^{-28}e.
\label{eqn:417}
\end{equation}
Therefore we have 
\begin{equation}
\det M^{(2)}\propto \lambda^{-20}e^{-3},
\label{eqn:418}
\end{equation}
where we have taken into account Eq.(\ref{eqn:324}) 
and used the fact 
$\Gamma^{(2)}B^{(2)-1}\Gamma^{(2)t}=0$.
The path-integral expression obtained is thus
\begin{equation}
Z^{(2)}=
\int{\cal D}g_{\mu\nu}{\cal D}b_{\mu}{\cal D}c^{\mu}
{\cal D}\bar{c}_{\mu}\Delta^{(2)}\exp\left(
i\int{\cal L}^{(2)}d^{4}x\right)
\label{eqn:419}
\end{equation}
with the measure
\begin{equation}
\Delta^{(2)}=(g^{00})^{4}(-g)^{-3/2}.
\label{eqn:420}
\end{equation}
This is in agreement with the result of 
Buchbinder-Lyahovich.\cite{BL86}

\section{Measures in more-than-four-derivative gravities}

We consider the system described by the Lagrangian 
\begin{eqnarray}
{\cal L}^{(n+2)} & = & 
{\cal L}_{\rm E}+{\cal L}_{\rm HD}^{(2)}
+{\cal L}_{\rm HD}^{(n+2)}+{\cal L}_{\rm BRS}^{(n+2)},
\label{eqn:501}\\
{\cal L}_{\rm HD}^{(n+2)} & = & 
\sqrt{-g}\left[ 
R^{\mu\nu}h_{2}\left(\frac{D^{2}}{\Lambda^{2}}\right) R_{\mu\nu}
+Rh_{0}\left(\frac{D^{2}}{\Lambda^{2}}\right) R
\right] .
\label{eqn:502}
\end{eqnarray}
For the BRS Lagrangian we take the following form: 
\begin{eqnarray}
{\cal L}_{\rm BRS}^{(n+2)} & = & 
\mbox{}-i\delta\left[
\bar{c}_{\mu}\omega\left(\frac{\Box}{\Lambda^{2}}\right)
\left(\frac{1}{\kappa}\partial_{\nu}\tilde{g}^{\mu\nu}
-\frac{\alpha}{2}\eta^{\mu\nu}b_{\nu}\right)\right]
\nonumber\\
& = & {\cal L}_{\rm GF}^{(n+2)}
+{\cal L}_{\rm FP}^{(n+2)},
\label{eqn:503}\\
{\cal L}_{\rm GF}^{(n+2)} & = & 
\frac{1}{\kappa}b_{\mu}\omega\left(
\frac{\Box}{\Lambda^{2}}\right)\partial_{\nu}\tilde{g}^{\mu\nu}
-\frac{\alpha}{2}\eta^{\mu\nu}b_{\mu}\omega\left(
\frac{\Box}{\Lambda^{2}}\right) b_{\nu},
\label{eqn:504}\\
{\cal L}_{\rm FP}^{(n+2)} & = & 
\mbox{}-\frac{i}{2}
(\partial_{\mu}\bar{c}_{\nu}+\partial_{\nu}\bar{c}_{\mu})
\omega\left(\frac{\Box}{\Lambda^{2}}\right)
D^{\mu\nu}_{\rho}c^{\rho}.
\label{eqn:505}
\end{eqnarray}
Here $h_{2}, h_{0}, \omega$ are polynomials of degree $n\ (>0)$ 
\begin{eqnarray}
h_{2}(x) & = & \sum_{k=0}^{n}\beta_{2k}x^{k},
\label{eqn:506}\\
h_{0}(x) & = & \sum_{k=0}^{n}\beta_{0k}x^{k},
\label{eqn:507}\\
\omega (x) & = & \sum_{k=0}^{n}\gamma_{k}x^{k};
\label{eqn:508}
\end{eqnarray}
$D^{2}\definition D^{\mu}D_{\mu}$ is
the covariant D'Alembertian associated with $g_{\mu\nu}$; 
$\Box\definition\partial^{\mu}\partial_{\mu}$ the ordinary 
D'Alembertian; and $\Lambda$ a dimentional constant.
In order to make the theory nonsingular we have introduced 
a polynomial $\omega (x)$ into the BRS Lagrangian.
The NL fields $b_{\mu}$ are treated as independent canonical 
coordinates here.
This is a point different from in the previous two sections, 
where $b_{\mu}$ were regarded as dependent fields to be 
integrated out in the course of path-integral calculation.

In the ADM coordinates the Lagrangians are written as follows:
\begin{eqnarray}
\lefteqn{\makebox[-2.8em]{}
{\cal L}_{\rm E}+
{\cal L}_{\rm HD}^{(2)}+
{\cal L}_{\rm HD}^{(n+2)}
}
\nonumber\\
& = & 
\left(\frac{1}{\Lambda\lambda}\right)^{2n}
\frac{1}{4\lambda^{3}}\sqrt{e}
\Lambda^{(n+2)ij,kl}
\nonumber\\
& & 
\mbox{}\times\left[
e_{ij}^{(n+2)}-\frac{\lambda^{(n+1)}}{\lambda}\left(
\dot{e}_{ij}-\nabla_{i}\lambda_{j}-\nabla_{j}\lambda_{i}\right)
-\left(\nabla_{i}\lambda_{j}^{(n+1)}
+\nabla_{j}\lambda_{i}^{(n+1)}
\right)\right]
\nonumber\\
& & 
\mbox{}\times\left[
e_{kl}^{(n+2)}-\frac{\lambda^{(n+1)}}{\lambda}\left(
\dot{e}_{kl}-\nabla_{k}\lambda_{l}-\nabla_{l}\lambda_{k}\right)
-\left(\nabla_{k}\lambda_{l}^{(n+1)}
+\nabla_{l}\lambda_{k}^{(n+1)}
\right)\right]
\nonumber\\
& & 
\mbox{}+\ldots ,
\label{eqn:509}\\
{\cal L}_{\rm GF}^{(n+2)}
& = & 
\gamma_{n}\left(\frac{1}{\Lambda\lambda}\right)^{2n}
\left[
\frac{1}{\kappa}\sqrt{e}\left(
\frac{1}{\lambda^{2}}\lambda^{(n+1)}b_{0}^{(n)}
-\frac{1}{\lambda^{2}}\lambda^{i}\lambda^{(n+1)}b_{i}^{(n)}
+\frac{1}{\lambda}e^{ij}\lambda_{i}^{(n+1)}b_{j}^{(n)}
\right)\right.
\nonumber\\
& & 
\makebox[5.5em]{}\!\left.\mbox{}\!
-\frac{\alpha}{2}\eta^{\mu\nu}b_{\mu}^{(n)}b_{\nu}^{(n)}
\right]
\nonumber\\
& & 
\mbox{}+\ldots ,
\label{eqn:510}\\
{\cal L}_{\rm FP}^{(n+2)}
& = & 
\gamma_{n}\left(\frac{1}{\Lambda\lambda}\right)^{2n}
i\sqrt{e}\left[
\frac{1}{\lambda}\bar{c}_{0}^{(n+1)}c^{0(n+1)}
+\frac{1}{\lambda^{2}}\bar{c}_{0}^{(n+1)}c^{0}
\lambda^{(n+1)}
+\frac{1}{\lambda}\bar{c}_{i}^{(n+1)}c^{i(n+1)}
\right.
\nonumber\\
& & 
\makebox[7em]{}\left.\mbox{}
+\frac{1}{\lambda}\bar{c}_{i}^{(n+1)}c^{0}\left(
\mbox{}-\frac{1}{\lambda}\lambda^{i}\lambda^{(n+1)}
+e^{ij}\lambda_{j}^{(n+1)}\right)\right]
\nonumber\\
& & 
\mbox{}+\ldots ,
\label{eqn:511}
\end{eqnarray}
where we have used the notation
\begin{equation}
\Lambda^{(n+2)ij,kl}\definition
\beta_{2n}P^{(2)ij,kl}+
4\left( \beta_{2n}+3\beta_{0n}\right) P^{(0)ij,kl}.
\label{eqn:512}
\end{equation}
We assume $\beta_{2n}\neq 0$ and $\beta_{2n}+3\beta_{0n}\neq 0$ 
for simplicity.

For the Hessian matrix 
\begin{equation}
M^{(n+2)} = 
\begin{array}{r@{}l}
  & \begin{array}{ccccccccc}
      \makebox[0.7em]{}e_{kl} & 
      \makebox[-0.5em]{}\lambda & 
      \makebox[-0.5em]{}\lambda_{b} & 
      \makebox[-0.5em]{}b_{\bar{0}} & 
      \makebox[-0.5em]{}b_{\bar{b}} & 
      \makebox[-0.5em]{}c^{0} & 
      \makebox[-0.5em]{}\bar{c}_{0} & 
      \makebox[-0.5em]{}c^{j} & 
      \makebox[-0.5em]{}\bar{c}_{j} 
    \end{array}\\
  \begin{array}{c}
    e_{ij} \\ \lambda \\ \lambda_{a} \\
    b_{\bar{0}} \\ b_{\bar{a}} \\
    c^{0} \\ \bar{c}_{0} \\
    c^{i} \\ \bar{c}_{i} 
  \end{array}
  & \left( 
      \begin{minipage}{13.5em}
      \begin{tabular}{ccccc|cccc}
        & & & &\makebox[-0.7em]{} &
            & &\makebox[-1em]{} & \\
        & & & &\makebox[-0.7em]{} &
            & &\makebox[-1em]{} & \\
        & & $A^{(n+2)}$ & &\makebox[-0.7em]{} &
            & $\Gamma^{(n+2)}$ &\makebox[-1em]{} & \\
        & & & &\makebox[-0.7em]{} &
            & &\makebox[-1em]{} & \\
        & & & &\makebox[-0.7em]{} &
            & &\makebox[-1em]{} & \\ \hline
        & & & &\makebox[-0.7em]{} &
            & &\makebox[-1em]{} & \\
        & & & &\makebox[-0.7em]{} &
            & &\makebox[-1em]{} & \\
        & & \raisebox{1.5ex}{$\Gamma^{(n+2)t}$} &
        &\makebox[-0.7em]{} &
            &\raisebox{1.5ex}{$B^{(n+2)}$} 
            &\makebox[-1em]{} & \\
        & & & &\makebox[-0.7em]{} &
            & &\makebox[-1em]{} &
      \end{tabular}
      \end{minipage}
    \right) ,
\end{array}
\label{eqn:513}
\end{equation}
we have the following form. 
The submatrix $A^{(n+2)}$ 
\begin{equation}
A^{(n+2)}=
\left(
 \begin{array}{ccccc}
 A^{(n+2)ij,kl} & A^{(n+2)ij,0} &
 A^{(n+2)ij,b} & A^{(n+2)ij,\bar{0}} &
 A^{(n+2)ij,\bar{b}}\\
 A^{(n+2)0,kl} & A^{(n+2)00} &
 A^{(n+2)0b} & A^{(n+2)0\bar{0}} &
 A^{(n+2)0\bar{b}}\\
 A^{(n+2)a,kl} & A^{(n+2)a0} &
 A^{(n+2)ab} & A^{(n+2)a\bar{0}} &
 A^{(n+2)a\bar{b}}\\
 A^{(n+2)\bar{0},kl} & A^{(n+2)\bar{0}0} &
 A^{(n+2)\bar{0}b} & A^{(n+2)\bar{0}\bar{0}} &
 A^{(n+2)\bar{0}\bar{b}}\\
 A^{(n+2)\bar{a},kl} & A^{(n+2)\bar{a}0} &
 A^{(n+2)\bar{a}b} & A^{(n+2)\bar{a}\bar{0}} &
 A^{(n+2)\bar{a}\bar{b}}
 \end{array}
\right)
\label{eqn:514}
\end{equation}
has the elements
\begin{eqnarray}
\makebox[-2em]{}
A^{(n+2)ij,kl} & = &
\left(\frac{1}{\Lambda\lambda}\right)^{2n}
\frac{\sqrt{e}}{2\lambda^{3}}
\Lambda^{(n+2)ij,kl},
\label{eqn:515}\\
\makebox[-2em]{}
A^{(n+2)ij,0} & = &
\left(\frac{1}{\Lambda\lambda}\right)^{2n}
\left( -\frac{\sqrt{e}}{2\lambda^{4}}\right)
\Lambda^{(n+2)ij,kl}
\left(
\dot{e}_{kl}-\nabla_{k}\lambda_{l}-\nabla_{l}\lambda_{k}\right) ,
\label{eqn:516}\\
\makebox[-2em]{}
A^{(n+2)ij,b} & = &
\left(\frac{1}{\Lambda\lambda}\right)^{2n}
\left( -\frac{\sqrt{e}}{2\lambda^{3}}\right)
\Lambda^{(n+2)ij,kl}
\left(
\nabla_{k}\delta_{l}^{b}+\nabla_{l}\delta_{k}^{b}\right) ,
\label{eqn:517}\\
\makebox[-2em]{}
A^{(n+2)ij,\bar{0}} & = & 0,
\label{eqn:518}\\
\makebox[-2em]{}
A^{(n+2)ij,\bar{b}} & = & 0,
\label{eqn:519}\\
\makebox[-2em]{}
A^{(n+2)00} & = &
\left(\frac{1}{\Lambda\lambda}\right)^{2n}
\frac{\sqrt{e}}{2\lambda^{5}}
\Lambda^{(n+2)ij,kl}
\nonumber\\
& & 
\makebox[5em]{}\times
\left(
\dot{e}_{ij}-\nabla_{i}\lambda_{j}-\nabla_{j}\lambda_{i}\right)
\left(
\dot{e}_{kl}-\nabla_{k}\lambda_{l}-\nabla_{l}\lambda_{k}\right) ,
\label{eqn:520}\\
\makebox[-2em]{}
A^{(n+2)0b} & = &
\left(\frac{1}{\Lambda\lambda}\right)^{2n}
\frac{\sqrt{e}}{2\lambda^{4}}
\Lambda^{(n+2)ij,kl}
\nonumber\\
& & 
\makebox[5em]{}\times
\left(
\dot{e}_{ij}-\nabla_{i}\lambda_{j}-\nabla_{j}\lambda_{i}\right)
\left(
\nabla_{k}\delta_{l}^{b}+\nabla_{l}\delta_{k}^{b}\right) ,
\label{eqn:521}\\
\makebox[-2em]{}
A^{(n+2)0\bar{0}} & = &
\left(\frac{1}{\Lambda\lambda}\right)^{2n}
\gamma_{n}\frac{1}{\kappa}\frac{\sqrt{e}}{\lambda^{2}},
\label{eqn:522}\\
\makebox[-2em]{}
A^{(n+2)0\bar{b}} & = &
\left(\frac{1}{\Lambda\lambda}\right)^{2n}
\gamma_{n}\left( -\frac{1}{\kappa}\right)
\frac{\sqrt{e}}{\lambda^{2}}\lambda^{\bar{b}},
\label{eqn:523}\\
\makebox[-2em]{}
A^{(n+2)ab} & = &
\left(
\stackrel{\leftarrow}{\nabla}_{i}\!\delta_{j}^{a}+
\stackrel{\leftarrow}{\nabla}_{j}\!\delta_{i}^{a}\right)
\left(\frac{1}{\Lambda\lambda}\right)^{2n}
\frac{\sqrt{e}}{2\lambda^{3}}
\Lambda^{(n+2)ij,kl}
\left(
\nabla_{k}\delta_{l}^{b}+\nabla_{l}\delta_{k}^{b}\right) ,
\label{eqn:524}\\
\makebox[-2em]{}
A^{(n+2)a\bar{0}} & = & 0,
\label{eqn:525}\\
\makebox[-2em]{}
A^{(n+2)a\bar{b}} & = &
\left(\frac{1}{\Lambda\lambda}\right)^{2n}
\gamma_{n}\frac{1}{\kappa}\frac{\sqrt{e}}{\lambda}e^{\bar{a}\bar{b}},
\label{eqn:526}\\
\makebox[-2em]{}
A^{(n+2)\bar{0}\bar{0}} & = &
\left(\frac{1}{\Lambda\lambda}\right)^{2n}
\gamma_{n}\alpha ,
\label{eqn:527}\\
\makebox[-2em]{}
A^{(n+2)\bar{0}\bar{b}} & = & 0,
\label{eqn:528}\\
\makebox[-2em]{}
A^{(n+2)\bar{a}\bar{b}} & = &
\left(\frac{1}{\Lambda\lambda}\right)^{2n}
\gamma_{n}\left( -\alpha\right)\eta^{\bar{a}\bar{b}}.
\label{eqn:529}
\end{eqnarray}
The submatrix $B^{(n+2)}$ is
\begin{equation}
B^{(n+2)}=
\left(\frac{1}{\Lambda\lambda}\right)^{2n}\gamma_{n}B^{(1)}.
\label{eqn:530}
\end{equation}

It turns out that the determinants are 
\begin{eqnarray}
\det A^{(n+2)} & \propto & \lambda^{-28n-28}e,
\label{eqn:531}\\
\det B^{(n+2)} & \propto & \lambda^{-16n-8}e^{4}.
\label{eqn:532}
\end{eqnarray}
Taking into account 
$\Gamma^{(n+2)}B^{(n+2)-1}\Gamma^{(n+2)t} =0$, 
we have
\begin{equation}
\det M^{(n+2)}\propto \lambda^{-12n-20}e^{-3}.
\label{eqn:533}
\end{equation}
Therefore the the measure in the path-integral 
\begin{equation}
Z^{(n+2)}=
\int{\cal D}g_{\mu\nu}{\cal D}b_{\mu}{\cal D}c^{\mu}
{\cal D}\bar{c}_{\mu}\Delta^{(n+2)}\exp\left(
i\int{\cal L}^{(n+2)}d^{4}x\right)
\label{eqn:534}
\end{equation}
is given as 
\begin{equation}
\Delta^{(n+2)}=(g^{00})^{3n+4}(-g)^{-3/2}.
\label{eqn:535}
\end{equation}

\section{Summary and discussion}

We have given a simple method of calculating path-integral measures 
in higher-derivative gravities. 
The measures are given as the generalized Lee-Yang terms. 
The results obtained in \S\S 3 and 4 agree with the ones of 
Buchbinder and Lyahovich,\cite{BL86}
while the result in \S 5 has been first reported 
in the present paper.

Our method proposed here is much simpler than
Buchbinder and Lyahovich's, which was 
to study the complex structure of the constraints 
of the theories and to perform canonical quantization 
straightforwardly.

For the four-derivative gravity their investigation covers more 
general cases than ours.
The constraints $\beta_{2}\neq 0$ 
and $\beta_{2}+3\beta_{0}\neq 0$ have been imposed 
for the parameters in \S 4.
In Ref~\citen{BL86} the cases of 
$(\beta_{2}=0, \beta_{2}+3\beta_{0}\neq 0)$ and
$(\beta_{2}\neq 0, \beta_{2}+3\beta_{0}=0)$ 
were also investigated. 
It is not difficult, however, 
to extend our method to treat those exceptional cases.
This will be discussed in a subsequent paper.\footnote{
A preliminary consideration on this subject has been given in 
Ref.~\citen{HN99}.
}

\section*{Acknowledgements}

Thanks are due to M.~Hirayama and T.~Kurimoto for discussion.


\begin{thebibliography}{99}
\bibitem{BL86} I.L.~Buchbinder and S.L.~Lyahovich, 
Class.~Quantum Grav. {\bf 4} (1987), 1487.

\bibitem{O50} M.~Ostrogradski, Mem.~Ac.~St.~Petersbourg {\bf V14} 
(1850), 385.

\bibitem{NH96} T.~Nakamura and S.~Hamamoto, \PTP{95,1996,469}.

\bibitem{HN99} S.~Hamamoto and M.~Nakamura, 
{\it Proceedings of the Ninth Workshop on General Relativity 
and Gravitation, Hiroshima, 1999}, ed. Y.~Eriguchi et al. 
(2000), p. 35.

\end{thebibliography}
\end{document}